\documentclass[reprint,twocolumn,aps,prb,reprint,superscriptaddress,longbibliography]{revtex4-2}
\usepackage{lipsum}
\usepackage{graphicx}
\usepackage{dcolumn}
\usepackage{bm}
\usepackage{color}
\usepackage{comment}
\usepackage{physics}
\usepackage{blindtext}
\usepackage{svg}
\usepackage{epsfig}
\usepackage{hyperref}
\hypersetup{
     colorlinks   = true,
     citecolor    = blue
}

\begin{document}

\title{Topological phase diagram of twisted bilayer graphene as a function of the twist angle}

\author{Leonardo A. Navarro-Labastida}
\affiliation{
 Posgrado en Ciencias Físicas, Instituto de F\'{i}sica, Universidad Nacional Aut\'{o}noma de M\'{e}xico (UNAM). Apdo. Postal 20-364, 01000 M\'{e}xico D.F., Mexico}
\author{Pierre A. Pantale\'on}
\email{pierre.pantaleon@imdea.org}
\affiliation{IMDEA Nanoscience, Faraday 9, 28049 Madrid, Spain}
\affiliation{Department of Physics, Colorado State University, Fort Collins, Colorado 80523, USA}
\author{Francisco Guinea}
\affiliation{IMDEA Nanoscience, Faraday 9, 28049 Madrid, Spain}
\affiliation{Donostia International Physics Center, Paseo Manuel de Lardiz\'abal 4, 20018 San Sebastián, Spain}
\author{Gerardo G. Naumis}
\email{naumis@fisica.unam.mx}
\affiliation{
 Depto. de Sistemas Complejos, Instituto de F\'{i}sica, Universidad Nacional Aut\'{o}noma de M\'{e}xico (UNAM). Apdo. Postal 20-364, 01000 M\'{e}xico, CDMX, Mexico
 }

\begin{abstract}
Twisted bilayer graphene (TBG) hosts a rich landscape of electronic phases arising from the interplay between strong electron-electron interactions and nontrivial band topology. While the flat bands near zero energy are central to many correlated phenomena, their interaction with higher-energy remote bands remains less understood. Here, we investigate these hybridization processes as a function of the twist angle and analyze their impact on the charge distribution, topological properties such as Chern number, quantum metric, and orbital magnetic energy. We identify multiple topological phase transitions between magic angles, driven by band inversions at high-symmetry points in momentum space. Notably, the central bands can exhibit phases with Chern numbers $C = \pm 2$, revealing previously unreported topological states in TBG.
 \end{abstract}

\maketitle 
\section{Introduction}

Twisted van der Waals (vdW) materials have emerged as a groundbreaking platform in condensed matter physics, offering new avenues for exploring strongly correlated and topological phases, as well as for developing next-generation quantum technologies. Conventional vdW materials are already known for their exceptional tunability, making them attractive for applications in electronic and optoelectronic devices~\cite{Sun2024Twisted,Kuang2024Optical}. A major challenge in harnessing their full potential lies in the inherently small lattice constant of two-dimensional materials, typically on the order of 1--2~\AA, which imposes strict constraints. For example, achieving sizable magnetic fluxes or strong correlation effects in such systems typically requires large magnetic fields or interaction strengths~\cite{Andrei2020Graphene,Balents2020Superconductivity}.

Twisted vdW systems, also known as moir\'{e} materials, overcome this limitation by introducing a twist angle between layers, giving rise to long-range moir\'{e} patterns. These patterns enlarge the effective unit cell to the nanometer scale---often hundreds of times larger than the atomic unit cell---thereby significantly reducing the required energy scales. This enhancement allows the emergence of a wide variety of strongly correlated phases, such as unconventional superconductivity, Mott insulators, and correlated Chern insulators~\cite{Cao2018,Yankowitz2019,Lu2019_SC_TBG,Stepanov2020_TBG,Oh2021evidence,Park2021_SC_TTG,Hao2021_SC_TTG,Kim2022evidence,Liu2022Isospin,Park2022_multi,Zhang2022_promotionSC}. Moreover, the moir\'{e} potential and twist angle provide a unique means to engineer flat bands with topologically nontrivial character, enabling the study of quantum geometry, Berry curvature effects, and emergent gauge fields. This tunability positions twisted vdW systems at the forefront of modern condensed matter research, serving as ideal platforms for investigating novel quantum phases and exploring the interplay between topology, strong correlations, and external fields~\cite{Carr2017,Andrei2020Graphene,Balents2020Superconductivity}.

An experimentally remarkable moir\'{e} system consists of two monolayer graphene sheets twisted at a specific angle, commonly referred to as TBG. This simple geometric modification gives rise to a rich variety of emergent quantum phases. Most notably, when the twist angle is tuned to a so-called "magic" value (approximately $1.1^\circ$)\cite{Bistritzer2011}, the electronic band structure exhibits nearly flat bands near the Fermi level, dramatically enhancing the role of electron-electron interactions~\cite{Cao2018}. As a result, unconventional superconductivity was observed in this system, alongside correlated insulating states and other symmetry-breaking phases~\cite{Cao2018,Yankowitz2019,Lu2019_SC_TBG,Stepanov2020_TBG,Oh2021evidence}. Remarkably, the phase diagram of TBG under gating and doping reveals striking similarities to that of high-temperature cuprate superconductors, including dome-shaped superconducting regions adjacent to correlated insulators~\cite{Armitage2010Progress}. These findings position TBG as a tunable, clean, and highly controllable platform for investigating the interplay between strong correlations, superconductivity, and topology in two-dimensional materials. Twisted moiré systems also offer a tunable setting for studying topological phase transitions~\cite{Carr2017}, where the Chern number of bands can be modified through changes in twist angle, electric field, or pressure. Such transitions are crucial for understanding the interplay between topology and strong correlations.

An open question in moir\'{e} physics concerns the hybridization of flat bands with higher-energy dispersive bands, which can lead to topological Lifshitz transitions~\cite{Volovik2017}—electronic reconstructions that fundamentally alter the band topology. This hybridization mechanism plays a critical role in defining the Chern number of moir\'{e} bands and influences key observables such as the Berry curvature, the quantum metric, and orbital magnetic moments. In TBG, the hybridization between the central flat bands and remote dispersive bands is one of the most challenging and consequential phenomena. It affects not only topological properties, but also the emergence of correlated insulating phases and unconventional superconductivity~\cite{Rhim2019,Datta2023Heavy}.

The interplay between band hybridization and electron-electron interactions determines whether TBG exhibits behavior characteristic of a Mott insulator, a topological Chern insulator, or a superconducting state~\cite{Jarillo-Herrero2018}. Although significant theoretical and experimental progress has been made, a number of fundamental questions remain unresolved. Chief among them is understanding how the flat bands interact with remote bands across different twist angles, and how these interactions influence topological phase transitions and the structure of the quantum geometry. Experimentally, this hybridization process remains elusive due to the difficulty in resolving the associated energy scales and wavefunction distributions.

In this work, we investigate the hybridization mechanism in TBG as a function of the twist angle. In Sec.~II, we present the theoretical framework within the chiral limit. In Sec.~III, we analyze the evolution of the electronic band structure and charge distribution. Sec.~IV explores topological phase transitions through changes in the Chern number and introduces a classification of the resulting phases. In Sec.~V, we examine the quantum geometry in regions of band inversion, and in Sec.~VI we analyze the orbital magnetic energy and its reorganization at high-symmetry points. We conclude in Sec.~VII, followed by technical details in the Appendices.

\section{Chiral TBG Hamiltonian}\label{secCTBG}

By stacking two rotated graphene layers with interlayer tunneling between nearest neighbors, it was demonstrated that narrow bands appear~\cite{Morell2011Electronic, Mele2012Band, Kindermann2011Landau, deGail2011Topologically, Bistritzer2011Moire, Mele2010Commensuration, TramblydeLaissardiere2010Localization, SuarezMorell2010Flat, LopesdosSantos2007Graphene}. In particular, a key feature of the models in Ref.~\cite{LopesdosSantos2007Graphene,Bistritzer2011Moire} is their formulation as a continuum model, incorporating both a gradient term and an interlayer coupling potential between carbon $\pi$ orbitals that varies smoothly with position. The specific form of this potential ensures that the interlayer hopping remains local and periodic, enabling Bloch-wave solutions for any small twist angle ($\theta < 3^\circ$).

Although those models contain the magic angle physics, in real TBG systems, the stacking points where carbon atoms are on top of another carbon (known as AA stacking regions), experience an out-of-plane displacement due to Coulomb repulsion, leading to a further reduction of the interlayer coupling~\cite{koshino2017}. As a result, the AB stacking regions, where a carbon atom in one layer sits at the center of a hexagon in the other layer, become larger than the AA regions. The interlayer tunneling associated with the stacking regions $AA$ and $AB$ are denoted as $w_{AA}$ and $w_{AB}$, respectively.

Due to these arguments and several experimental observations, Tarnopolsky et.al~\cite{Tarnpolsky2019,Ledwith2021} neglected the interlayer electron tunneling in AA regions, i.e., $w_{AA}=0$. This led to the so-called chiral model Hamiltonian. To write such a model, consider as a basis the wave function $\Psi(r)=\begin{pmatrix} 
\psi_1(r) ,
\psi_2(r),
\chi_1(r),
\chi_2(r)
\end{pmatrix}^T$ where $\psi_j(r)$ and $\chi_j(r)$ are the Wannier orbitals on each site of the graphene's unit cell, i.e., at sites A and B of the graphene's bipartite lattice. The subscripts $1,2$ represent each graphene layer. The chiral Hamiltonian in this basis is \cite{Tarnpolsky2019,Khalaff2019, Ledwidth2020}, 
\begin{equation}
\begin{split}
\mathcal{H}
&=\begin{pmatrix} 
0 & D^{\ast}(-r)\\
 D(r) & 0
  \end{pmatrix}  \\
\end{split} 
\label{H_initial}
\end{equation}
where the zero mode operator is defined as, 
\begin{equation}
\begin{split}
D(r)&=\begin{pmatrix} 
-i\Bar{\partial} & \alpha U(r)\\
  \alpha U(-r) & -i\Bar{\partial} 
  \end{pmatrix}  \\
\end{split} 
\end{equation}
with
$\Bar{\partial}=\partial_x+i\partial_y$ is the antiholonomic derivative and $\partial=\partial_x-i\partial_y$ the holonomic one.  The distance between Dirac cones in a moiré valley is $k_{\theta}=2k_{D}\sin{\frac{\theta}{2}}$ with $k_{D}=\frac{4\pi}{3a_{0}}$ the Dirac wave vector and $a_{0}$ the lattice constant of graphene. The physics of this model is solely captured via the parameter $\alpha$, defined as,
\begin{equation}\label{eq:alphaa}
    \alpha=\frac{w_{AB}}{v_0 k_\theta}
\end{equation}
with value $w_{AB}=110$ meV and $v_0$ is the Fermi velocity $v_0=\frac{19.81eV}{2k_D}$. For  TBG the interlayer potential is \cite{Tarnpolsky2019},

\begin{equation}
    U(\bm{r})=\sum^{3}_{\mu=1}e^{-i(\mu-1)\phi}e^{-i\bm{q}_{\mu}\cdot \bm{r}}
\end{equation}
with $\phi=2\pi/3$ and, 
\begin{equation}
    \bm{q}_{\mu}=k_{\theta}(\sin{[(\mu-1)\phi]},-\cos{[(\mu-1)\phi]}).
\end{equation}
The vectors $\bm{b}_{1,2}=\bm{q}_{2,3}-\bm{q}_{1}$ turn out to be the moir\'e reciprocal vectors and for its utility, we define a third vector $\bm{b}_{3}=\bm{q}_{3}-\bm{q}_2$.

\begin{figure*}[htp]
\centering
\includegraphics[width= 0.95\textwidth]{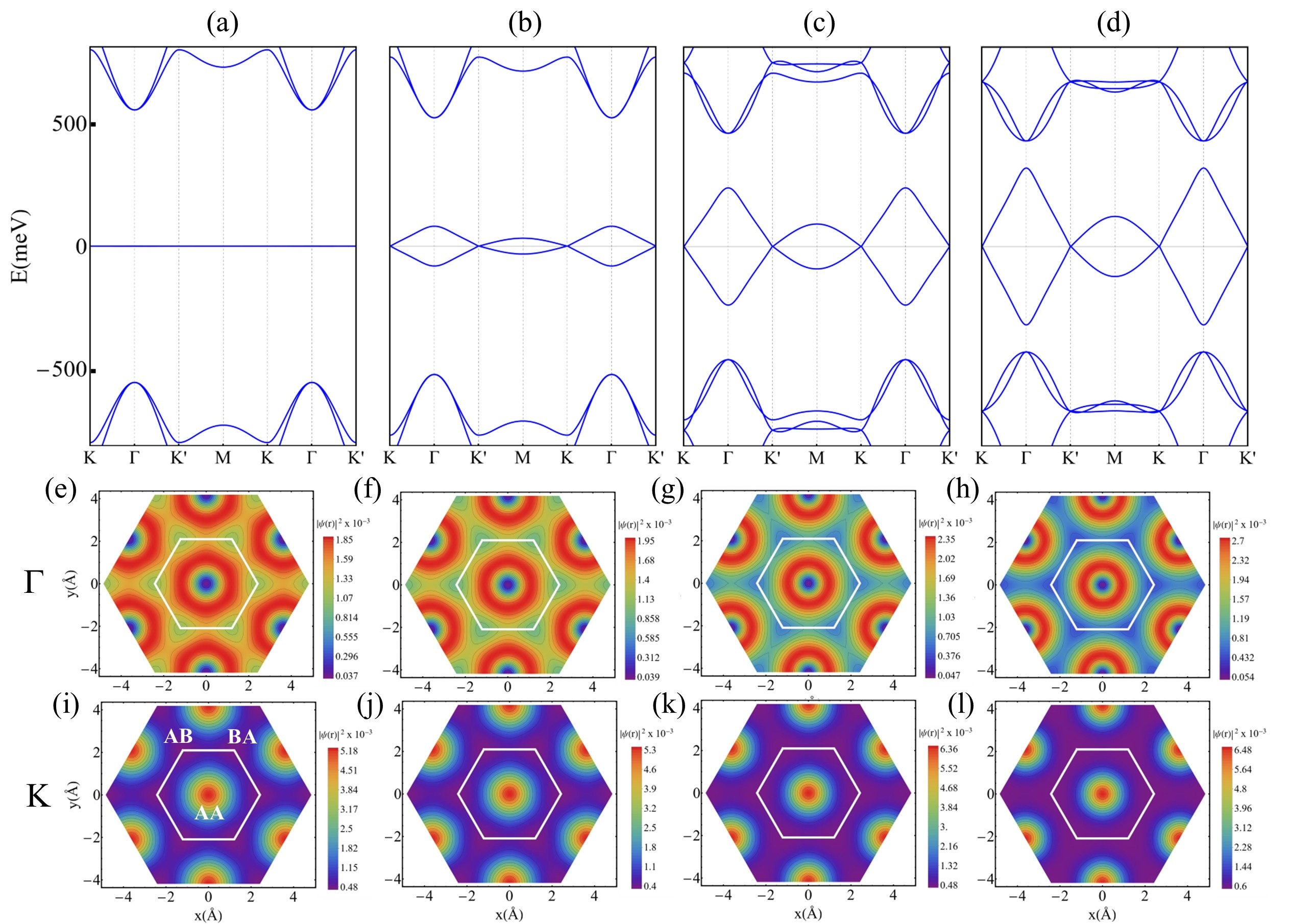}
\caption{Evolution of the band structure and charge density for (a) $\alpha = 0.586$, (b) $\alpha = 0.65$, (c) $\alpha = 0.80$, and (d) $\alpha = 0.90$. Panels (e)–(h) show the corresponding charge distributions for the same values of $\alpha$, evaluated at the $\Gamma$ point (middle row). Panels (i) and (j) show the charge distributions at the $K$ point (bottom row). The AA, AB, and BA stack configurations of TBG are shown in (i).}
\label{fig:carga}
\end{figure*}

Here we will use units where $v_0=1$ and renormalized vectors such that   $\bm{q}_{\mu} \rightarrow \frac{\bm{q}_{\mu}}{k_{\theta}}$ and $\bm{r} \rightarrow k_{\theta}\bm{r}$. In other words, the system can be viewed as if it has fixed geometry, i.e., setting, $\bm{q}_{1}:=(0,-1),\bm{q}_{2}:=(\sqrt{3}/2, 1/2),
\bm{q}_{3}:=(-\sqrt{3}/2, 1/2)$,
while the twist angle enters only in the coupling parameter $\alpha$. It is also convenient to define a set of unit vectors $\bm{q}_{\nu}^{\perp}$ perpendicular to the set $\bm{q}_{\nu}$ and defined as $\bm{q}_{1}^{\perp}=(1,0),\bm{q}_{2}^{\perp}=\big(-\frac{1}{2},\frac{\sqrt{3}}{2}\big),\bm{q}_{3}^{\perp}=\big(-\frac{1}{2},-\frac{ \sqrt{3}}{2}\big)$.

The moir\'e vectors  unit cell are given by $\bm{a}_{1,2}=(4\pi/3k_{\theta})(\sqrt{3}/2,1/2)$. Note that $\bm{q}_{\nu}\cdot \bm{a}_{1,2}=-\phi$ for $\nu=1,2,3$.
As explained in the appendix \ref{sec:squared}, the square of the Hamiltonian ($H^{2}$) allows to transform the problem into an electron moving under the action of a non-abelian field \cite{Guinea2012,Naumis2023r,NN2024}. As shown in Sec. \ref{Sec:Non-Abelian}, the Hamiltonian  $H^{2}$ allows to understand how the problem renormalizes once $\alpha \rightarrow \infty$, therefore leading to an effective abelian quantum Hall effect~\cite{NN2024}. 

\begin{figure*}[ht]
\centering
\includegraphics[scale=0.75]{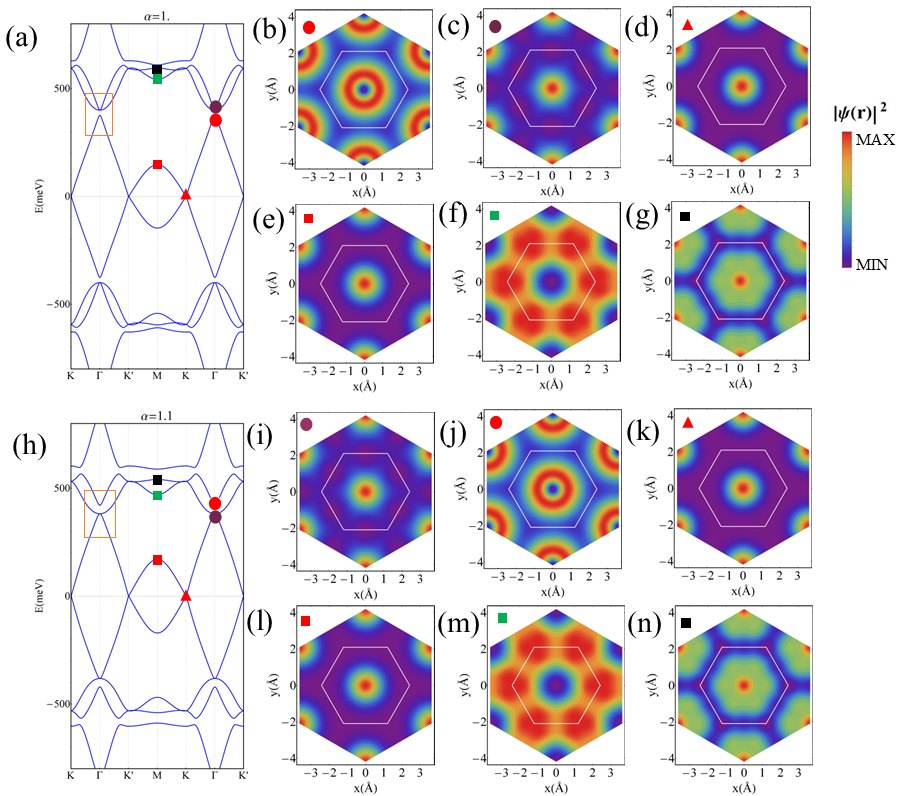}
\caption{Band structure for (a) $\alpha=1.0$ and (h) $\alpha=1.1$. Density plots in all panels display the charge density at the high symmetry points indicated with colored shapes in (a) and (h), respectively.\ White hexagon is the real space unit cell.\ Orange rectangle indicates the regions where a band inversion occurs.}
\label{fig:alpha1}
\end{figure*}

\begin{figure*}[ht]
\centering
\includegraphics[scale=0.75]{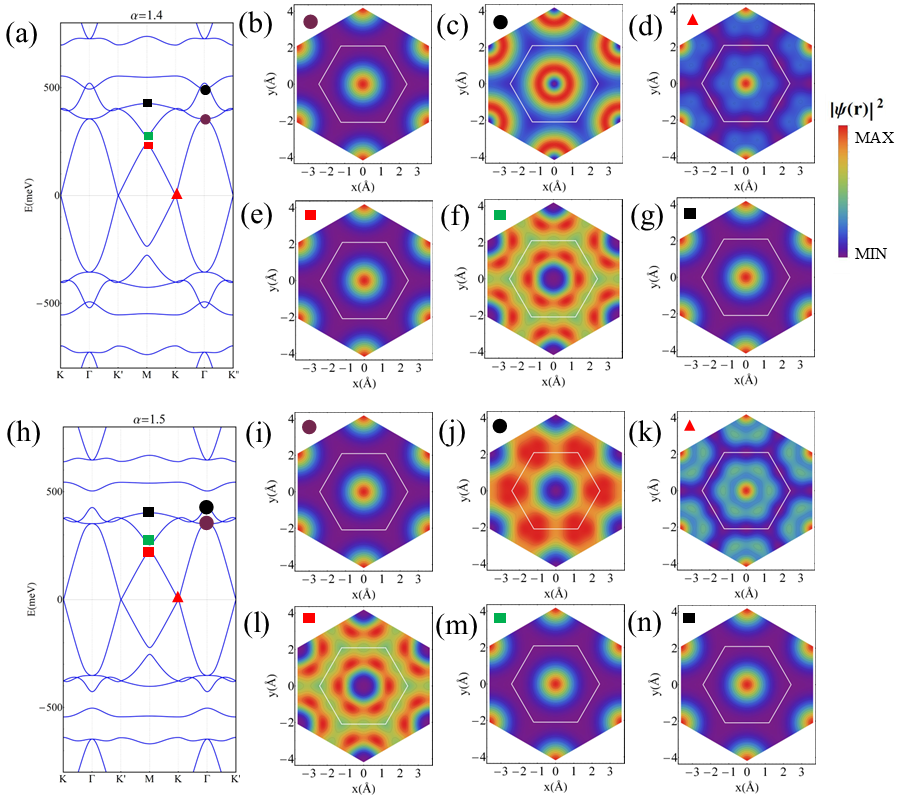}
\caption{Band structure for (a) $\alpha=1.4$ and (h) $\alpha=1.5$. Density plots in all panels display the charge density at the high symmetry points indicated with colored shapes in (a) and (h), respectively. The white hexagon is the real space unit cell.}
\label{fig:alpha2}
\end{figure*}

The electronic spectrum can be found from the Schrödinger equation $\mathcal{H} \Psi_{\bm{k}}(\bm{r})=E_{\bm{k}} \Psi_{\bm{k}}(\bm{r})$ for different angles using Bloch theory for a given reciprocal vector $\mathbf{k}$, where $\Psi_{\bm{k}}(\bm{r})$ is a bispinor. Some examples are presented in Fig. \ref{fig:alpha1} and Fig.~\ref{fig:alpha2}. As is well documented, flat bands arise in this system for certain magic values of $\alpha=\alpha_m$, where $m=1,2,3,...$ is the order of these magic angles~\cite{Tarnpolsky2019}. These flat-band modes have zero dispersion, $E_{\bm{k}}=0$, for all ${\bm{k}}$ and thus are also known as zero modes. In the chiral TBG, these zero modes follow the recurrence relation $\alpha_{n+1}-\alpha_{n} \approx 3/2$~\cite{Naumis2023,2024NL,Naumis2021,Naumis2022}  where the first magic angles given by $\alpha_1=0.586$, $\alpha_2=2.221$, $\alpha_3=3.751$, $\alpha_4=5.276$, $\alpha_5=6.795$, $\alpha_6=8.313$, $\alpha_7=9.829$, $\alpha_8=11.345$, and so forth. 

\section{Electronic Spectrum and Charge Density Between Magic Angles}\label{ChargeDistribution}

As the twist angle is reduced from the first magic angle, the two middle bands become more dispersive \cite{MacDonald2011}. This behavior is illustrated in the top row of Fig.~\ref{fig:carga}. Therein we show the evolution of the band structure from the first magic angle, $\alpha_1 = 0.586$, to $\alpha = 0.90$, where the bands are strongly dispersive. In the middle and bottom rows of Fig.~\ref{fig:carga}, we show the charge distributions at the $K$ and $\Gamma$ points, respectively. At the magic angle, shown in Fig.~\ref{fig:carga}~(e), the charge density at the $\Gamma$ point forms a ring-like pattern around the AA regions (colored in red), with slightly lower density near the AB and BA sites (colored in yellow/green).\ The charge density reaches a minimum at the AA center.\ In contrast, at the $K$ point, shown in Fig.~\ref{fig:carga}(i), the charge is concentrated at the AA center, with additional contributions from the regions between the AB and BA domains. As the twist angle decreases (by increasing $\alpha$), a redistribution of charge is observed, as seen by following the panels from Fig.~\ref{fig:carga}(e) to Fig.~\ref{fig:carga}(h), and from Fig.~\ref{fig:carga}(i) to Fig.~\ref{fig:carga}(l).\ At the $K$ point, the charge becomes even more localized at the AA center, while at the $\Gamma$ point, the ring-like pattern persists but exhibits a depletion of charge from the AB and BA regions.

\subsection{\texorpdfstring{$\Gamma$ Point Transition}{Gamma Point Transition}}

By further reducing the twist angle, we observe an intriguing behavior in the electronic properties.\ Figure~\ref{fig:alpha1}(a) shows the electronic band structure for $\alpha = 1.0$.\ At the $\Gamma$ point (highlighted by the orange rectangle), the first two remote bands are degenerate.\ The corresponding charge distribution, shown in Fig.~\ref{fig:alpha1}(c), is concentrated at the AA sites (red regions), with minimal charge density (purple regions) located between the AB and BA domains, forming a kagome-like pattern.\ In Fig.~\ref{fig:alpha1}(h), for $\alpha = 1.1$, a band inversion is evident.\ Here, the upper middle band becomes degenerate with the first remote band, indicating a transition in which the original degeneracy at the $\Gamma$ point is exchanged.\ The resulting charge distribution is shown in Fig.~\ref{fig:alpha1}(i) for the degenerate middle and first remote band, and in Fig.~\ref{fig:alpha1}(j) for the second remote band.\ We also show the charge distributions at other high-symmetry points, where the spatial profile of the charge remains unchanged. We note that the degeneracy between the middle and first remote band at $\Gamma$ is persistent in a range of $\alpha$ values (see Sec. IV).\ On the other hand, it is known that near the first magic angle, the electronic structure of TBG can be characterized by a set of localized flat-band orbitals centered at the AA-stacking regions, and a set of extended bands that are energetically separated from these localized states~\cite{Song2022Magic, Haule2019Strongly, Datta2023Heavy, Shi2022Heavy, Kang2021Cascades}.\ This distinction is illustrated in Fig.~\ref{fig:alpha1}(d) and Fig.~\ref{fig:alpha1}(e), where the charge is localized at the AA centers, and in Fig.~\ref{fig:alpha1}(b), where the charge is highly concentrated in a ring-like pattern.\ However, by inspecting the charge distribution for $\alpha$ values beyond the transition, Fig.~\ref{fig:alpha1}(i) to Fig.~\ref{fig:alpha1}(n), it becomes clear that the exchange of degeneracy at $\Gamma$ results in an electronic structure that is different from that resulting of a combination of localized flat-band orbitals and nearly-free conduction states~\cite{Shi2022Heavy,Song2022Magic}. 

\begin{figure*}[ht]
\centering
\includegraphics[scale=0.75]{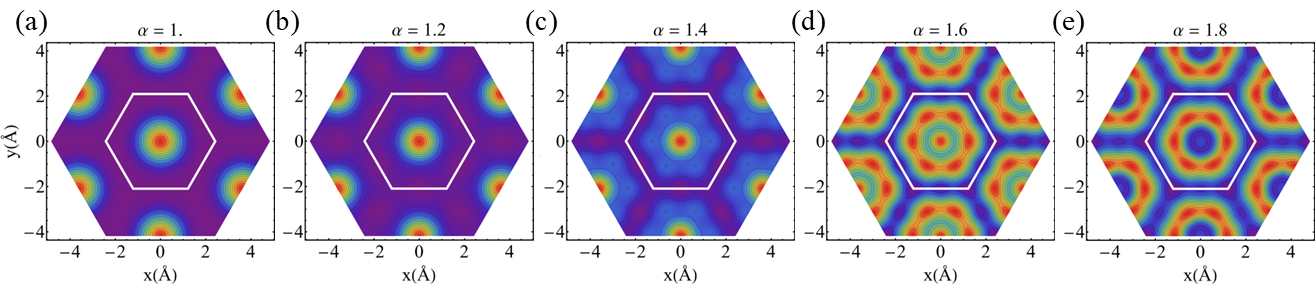}
\caption{Charge density at the $K$ point as a function of the twist angle. The color scale indicates charge density, with purple representing the minimum and red the maximum values.}
\label{fig:kpoint}
\end{figure*}

\subsection{M Point Transition}

As the $\alpha$ parameter increases, the twist angle is further reduced. We find that within a certain range of $\alpha$, the degeneracy at the $\Gamma$ point persists (see Sec. IV). However, within this degenerate region, an additional transition occurs. As shown in Fig.~\ref{fig:alpha2}(a) and Fig.~\ref{fig:alpha2}(h), a band inversion takes place at the $M$ point. The corresponding charge distributions before and after the transition are displayed in Fig.~\ref{fig:alpha2}(e)–(f) and Fig.~\ref{fig:alpha2}(l)–(m), respectively. Notably, although the Dirac points $\bm{K}$ and $\bm{K}^{\prime}$ are not directly involved in the band inversion, the charge distributions at these points no longer resemble those near the first magic angle after the transition. At the $M$ point, the charge distribution forms puddles arranged in a hexagonal ring around the AA centers, where the charge density is minimal. A similar pattern is observed at the $K$ point; however, in this case, the charge density is maximal at the AA centers.

\subsection{Nonlocal Effects of Band Transitions on Charge Density}

As mentioned earlier, although the Dirac points $\bm{K}$ and $\bm{K}^{\prime}$ are not directly involved in the band inversion shown in Fig.~\ref{fig:alpha2}, their charge distributions are nonetheless significantly affected. This modification arises immediately after the first transition described in Fig.~\ref{fig:alpha1}, and is illustrated in Fig.~\ref{fig:kpoint}. It is evident that as $\alpha$ increases, charge puddles begin to form an emerging hexagonal structure around the AA centers. This behavior can be attributed to the changes in the moiré potential induced by the decreasing twist angle. As $\alpha$ increases, it has been shown that the combined effects of moiré confinement and the quantized orbital motion of electrons effectively create a quantum well~\cite{NN2024}. The quantized orbital motion generates a centrifugal-like potential, analogous to the behavior of angular momentum in a two-dimensional quantum harmonic oscillator. This effect leads to a redistribution of charge centered at the $\Gamma$ point. As $\alpha$ continues to increase, this redistribution extends throughout the entire mBZ, ultimately altering the spatial profile of the electronic states, including those at the Dirac points.

\section{Topological phase transitions}\label{Topologicalphases}

Twisted bilayer graphene has been shown to host Landau levels~\cite{CANO2021, JieWang2023, Ledwidth2023, patrickL2023}, which play a central role in its remarkable electronic properties~\cite{KhalafEslam2021, Jarrillo2021_Fractional, 2021yardenn, 2021ShangL, 2022Vishwanath, Ledwith2021, 2022Ledwith_Vortex, 2022Ashvin}. Several previous studies have explored the interplay between flat band touchings and the emergence of correlated phases such as superconductivity. For instance, Refs.~\cite{kukka2022, 2025Bernevig} investigate toy models where flat bands and dispersive bands intersect at high-symmetry points, leading to an enhancement of the quantum geometric tensor, particularly the quantum metric.

In this work, we reveal the existence of new topological phases arising in the hybridization regions where the central flat band touches higher-energy bands.\ These touchings often occur at high-symmetry points such as $\Gamma$ and $M$, and are accompanied by topological transitions characterized by changes in the Chern number.\ To analyze these phases, we begin by calculating the Berry curvature using the formalism of Ref.~\cite{Vanderbilt1997}, allowing us to track the evolution of topological invariants across different hybridization regimes. We write, $\bm{\Omega}(\bm{k})=\nabla_{k}\times\mathcal{\bm{A}}(\bm{k})$ where $\mathcal{\bm{A}}(\bm{k})= i\langle\Psi(\bm{k})|\nabla_{\bm{k}}|\Psi(\bm{k})\rangle$ is the Berry connection.\ The Chern number $C$ is given by

\begin{equation}
\begin{split}\label{eq:BerryConnector2}
\int_{S}\bm{\Omega}\cdot d\bm{S} = 2\pi C. 
\end{split}
\end{equation}
To numerically solve the above equation, we introduced an small staggered sublattice potential in the diagonal of the Hamiltonian Eq.~(\ref{H_initial}), i.e., 
\begin{equation}
\begin{split}\label{eq:staggered}
V_{\delta}=\frac{\delta}{2}\sigma_z,
\end{split}
\end{equation}
with $\delta=10$ meV, in this way, trivial singularities for the band touching are removed. To calculate the Chern number we use the numerical method described in Ref.~\cite{Vanderbilt1997,Fukui2005Chern}.

\begin{figure*}[tb]
\centering
\includegraphics[scale=0.48]{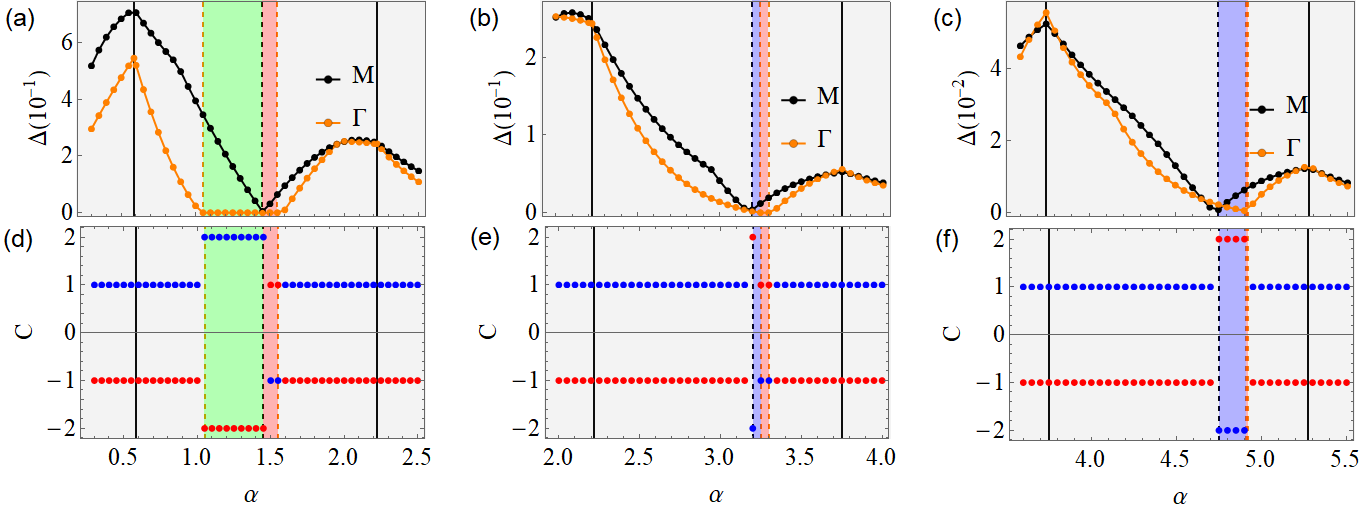}
\caption{The upper row (a)–(c) shows the energy gap as a function of $\alpha$ for the first three hybridization processes, while the lower row (d)–(f) presents the corresponding Chern numbers as a function of $\alpha$ for the same processes. In the energy gap plots, we include both $\bm{k} = \bm{\Gamma}$ and $\bm{k} = \bm{M}$ points, as these high-symmetry points appear to play a central role in the hybridization mechanism and are directly associated with the topological phase transitions. Notably, only the first hybridization process differs significantly from the higher-order processes, particularly in the sequence of gap closings. The color coding represents different topological phases as described in the main text.}
\label{fig:chern_vs_gap} 
\end{figure*}

In Fig.~\ref{fig:chern_vs_gap} we show the Chern numbers for the two middle bands as a function of $\alpha$. The topological phase at magic angles (indicated by vertical black lines) has Chern numbers $\{C_v, C_c\} = \{-1,1\}$ for the valence (red) and conduction band (blue), in agreement with previous works~\cite{Tommaso2020,Ledwith2021,Long2023Electronic, Repellin2020Frac}. We denote as $P_1$, $P_2$, $P_3$ and $P_4$ the topological phases with corresponding Chern numbers $\{-1,1\}$ (gray regions), $\{-2,2\}$ (green regions), $\{2,-2\}$ (blue regions) and $\{1,-1\}$ (red regions). 

Phase $P_1$ is present across all magic angles, indicating a robust topological regime. In contrast, phase $P_2$ emerges specifically between the first and second magic angles, and $P_4$ between the first and third magic angles, highlighting regions of repeated band hybridization. Finally, phase $P_3$ appears only in the high-$\alpha$ regime, beyond the second magic angle. We note that flat bands at magic angles are always in phase $P_1$~\cite{2021Ledwith}. 

In Fig.~\ref{fig:chern_vs_gap}, we also show the energy gaps at the $\bm{\Gamma}$ and $\bm{M}$ points between the central band and the first remote band. A clear correlation is observed between the closing of these gaps and the corresponding jumps in the Chern number of the central band. Notably, the variation in the Chern number is more pronounced at the $\bm{M}$ point than at $\bm{\Gamma}$. This is because there are three inequivalent $\bm{M}$ points in the mBZ, and a band inversion at these points leads to the transfer of topological charge between adjacent bands at each of them. This multi-point charge transfer amplifies the total change in the Chern number and is illustrated in Fig.~\ref{fig:alpha2} and Fig.~\ref{fig:alpha1}. Furthermore, the sequence of Chern number transitions differs between the $\bm{M}$ and $\bm{\Gamma}$ points, reflecting distinct gap-closing mechanisms and topological charge redistribution at these high-symmetry points, which ultimately shape the topological character of the emerging phases.

\begin{table}[h!]
\centering
\begin{tabular}{|>{\centering\arraybackslash}p{1cm}|>{\centering\arraybackslash}p{2cm}|>{\centering\arraybackslash}p{1.5cm}|}
\multicolumn{3}{c}{} \\
\hline
Phase & $\alpha$ interval & $[C_v, C_c]$ \\
\hline
$P_2$ & $[1.030, 1.500]$ & $[-2, 2]$ \\
$P_4$ & $[1.501, 1.650]$ & $[1, -1]$ \\
\hline
$P_3$ & $[3.180, 3.250]$ & $[2, -2]$ \\
$P_4$ & $[3.260, 3.310]$ & $[1, -1]$ \\
\hline
$P_3$ & $[4.730, 4.913]$ & $[2, -2]$ \\
$P_4$ & $[4.914, 4.915]$ & $[1, -1]$ \\
\hline
\end{tabular}
\caption{Topological phases at the first three hybridization regions. The intervals in $\alpha$ where they are found are indicated as well as the Chern number of the central bands.}
\label{tab:topo_phases}
\end{table}

Table I summarizes the newly found topological phases  at the indicated $\alpha$ interval. Each of them is characterized by a certain band hybridization process. Something to emphasize here is that Phase $P_2$ only exists in between the first and second magic angles, while phase $P_3$ appears in the second hybridization region. 
According to Table~\ref{tab:topo_phases}, the first hybridization region, which occurs between the first magic angle $\alpha_1 = 0.586$ and the second magic angle $\alpha_2 = 2.221$, is notably different from the higher hybridization regions, where only phases $P_3$ and $P_4$ are observed. In contrast, this first region exhibits both phases $P_2$ and $P_4$, indicating a richer topological structure. 

As shown in Fig.~\ref{fig:alpha1}, the first gap closing occurs at $\alpha = 1.030$ and is associated with the $\bm{\Gamma}$ point. The bands remain hybridized over a continuous range of $\alpha$, and a second gap closing takes place at the $\bm{M}$ point around $\alpha = 1.45$. However, the gap at $\bm{\Gamma}$ remains closed until approximately $\alpha = 1.650$, consistent with the interval of phase $P_4$ reported in Table~\ref{tab:topo_phases}. Furthermore, Table~\ref{tab:topo_phases} reveals that phase $P_2$ appears only in the first hybridization region and is absent in the higher-order ones. This suggests that as the magic angle index increases, the appearance of phase $P_2$ diminishes, while phase $P_3$ becomes more prominent. From these observations, we conclude that for higher values of $\alpha$, the topology of the system is primarily governed by the gap closing at the $\bm{M}$ point. In general, we identify two dominant hybridization sequences, which are summarized in Table~II.

\begin{table}[h!]
\centering
\begin{tabular}{p{2cm}|p{3.3cm}}

\multicolumn{2}{c}{} \\
\hline
First hybridization& $\bm{\Gamma} \rightarrow\bm{M}\rightarrow\bm{\Gamma} $\\
\hline
$\theta$ -twist angle & $0.60^{\circ}\rightarrow0.43^{\circ}\rightarrow0.40^{\circ}$ \\
\hline
Higher hybridizations& $\bm{M}\rightarrow\bm{\Gamma}\rightarrow\bm{\Gamma}$ \\
\hline
\end{tabular}
\caption{Gap closing sequence order in hybridization regions and experimental twist angles for the first hybridization region. These are the numerical values as a reference for the twist angle where it is expected to find these topological regions; a bias voltage between layers has to be introduced $V_{\delta}=10 \text{ meV}$.}
\label{tab:sequence}
\end{table}

Table~\ref{tab:sequence} also includes the estimated experimental twist angles at which the first-order hybridization process occurs. In contrast, higher-order hybridization processes are more difficult to realize experimentally, as they require extremely small twist angles. Achieving such precise control remains technically challenging with current fabrication techniques.

Notice that the existence of phase $P_2$ can be traced back to a key finding demonstrated in Refs.~\cite{Naumis2022,Naumis2023,Naumis2023r}. Specifically, the first magic angle region is distinct because the system behaves as if it were subjected to a non-Abelian magnetic field. In contrast, for larger twist angles, the Schrödinger equation enters the boundary layer limit, resulting in an effective Abelian magnetic field. This transition in behavior is further discussed and summarized in Sec.~\ref{Sec:Non-Abelian}.

\section{Quantum Geometry at Inversion Points}\label{QuantumM}


We found that the change in the Chern number during a topological phase transition is associated with a band inversion at high-symmetry points, followed by an exchange of the charge distribution between the bands involved in the transition~\cite{Fu2027Topological}.  

\begin{figure*}[tb]
\centering{
\includegraphics[scale=0.50]{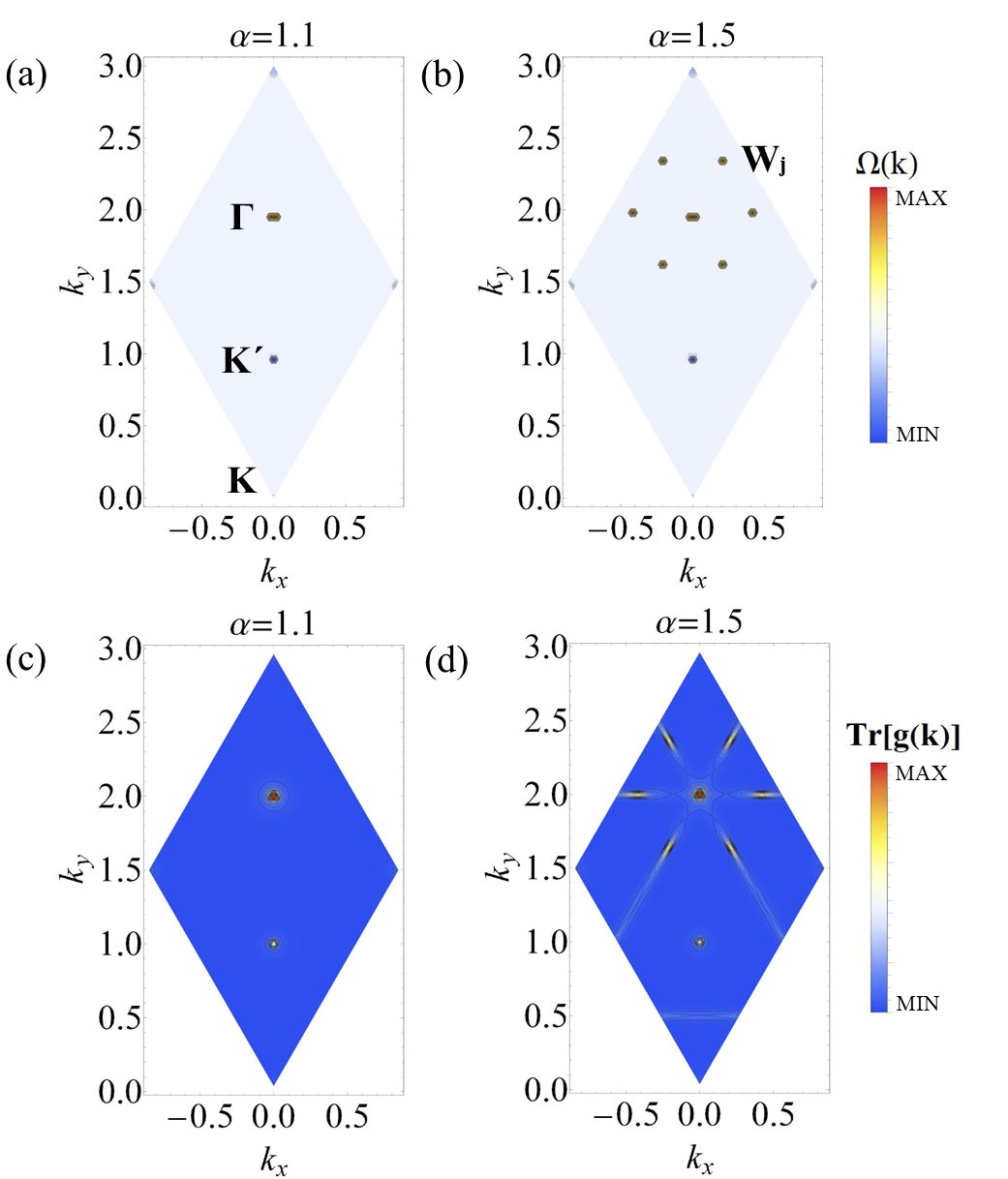}}
\caption{Berry curvature for central bands at the hybridization regions. The Berry curvature has a strong density at $\bm{\Gamma}$, $\bm{K}$, $\bm{K}'$ and $\bm{W}_j$ points. This unit cell is defined by the moiré reciprocal vectors $\bm{b}_{1,2}$ that form a parallelogram. All the corners are $\bm{K}$ Dirac points with the origin defined at $\bm{K}=(0,0)$. In (a) $\alpha=1.1$ a gap is close to the $\bm{\Gamma}$ point, and then exists a Berry curvature around such points of $\pm 1$ for the conduction (valence) central band. In (b), when the gap closes and reopens in $\bm{M}$ at $\alpha=1.5$, this produces an extra Berry curvature of $\pm 1/2$ around the $\bm{W}_j$ points over the conduction (valence) central bands. Central band quantum metric as a function of momentum at hybridization regions for (c) $\alpha=1.1$ and (d) $\alpha=1.5$. At the $\bm{\Gamma}$ point, i.e., $\bm{k}=-2\bm{q}_{1}$ the quantum metric is maximum for both cases while for the case $\alpha=1.5$ there is an extra density in $6$ spots around $\bm{W}_j$ points. For these numerical results, we considered the second band $n=2$.}
\label{fig:berry}
\end{figure*}

Figure~\ref{fig:berry} shows the Berry curvature $\Omega(\bm{k})$ for the upper central band at values of $\alpha$ where band inversion occurs. Due to the particle-hole symmetry of the system, a similar behavior is observed in the lower central band. At the first transition, shown in Fig.~\ref{fig:berry}a), the Berry curvature is strongly concentrated around the $\bm{\Gamma}$, $\bm{K}$, and $\bm{K}'$ points. The large contribution at $\bm{\Gamma}$ arises from its proximity to the remote bands (see also Fig.~\ref{fig:alpha1}), while the peaks at $\bm{K}$ and $\bm{K}'$ result from the small staggered sublattice potential introduced in the model. In Fig.~\ref{fig:berry}b), the Berry curvature remains concentrated at these same points, with additional significant contributions from the six points surrounding $\bm{\Gamma}$, labeled as $\bm{W}_j$ with $j = 1, 2, \dots, 6$.

We now examine the points where the Berry curvature is concentrated. To quantify the local flux contribution, we compute

\begin{equation}
\mathcal{C}_n(\mathbf{k}) = \frac{1}{2\pi} \mathbf{S}{g}(\mathbf{k}) \cdot \mathbf{\Omega}_n(\mathbf{k}),
\end{equation}
where $n$ is the band index and $\mathbf{S}_g(\mathbf{k})$ is a small grid plaquette enclosing the point $\mathbf{k}$ within a distance proportional to $1/\delta$. We find that for all values of $\alpha$, the points $\bm{K}$ and $\bm{K}'$ consistently contribute fluxes of $1/2$ and $-1/2$, respectively, as expected from the linear dispersion near the Dirac points~\cite{Berry1984Quantal}.

The notable changes occur at the band inversion points, particularly at $\bm{\Gamma}$ and $\bm{M}$. When the inversion takes place at $\bm{M}$, the six surrounding points $\bm{W}_j$ acquire non-zero Berry curvature. In the central band (indexed as $n = 0$ due to chiral symmetry), we find $\mathcal{C}_0(\bm{W}j) = -1/2$, while for the first remote band ($n = 1$), $\mathcal{C}_1(\bm{W}_j) = +1/2$. For the second remote band ($n = 2$), the contribution vanishes: $\mathcal{C}_2(\bm{W}_j) = 0$. We define the total Berry flux due to these points as,
\begin{equation}
\begin{split}\label{eq:berryMpoint}
\mathcal{C}^{\bm{W}}_n=\sum^{6}_{j=1}\mathcal{C}_{n}(\bm{W}_j)
\end{split}
\end{equation}
For the central band, $\mathcal{C}^{\bm{W}}_{0} \approx 6(\pm 1/2)=  +3$, while for the first remote band $\mathcal{C}^{\bm{W}}_{1} \approx 6(\pm -1/2)=-3$ and for the second remote band $\mathcal{C}^{\bm{W}}_{2} \approx 0$ . It follows a sum rule for the flux of bands,
\begin{equation}
\begin{split}\label{eq:berryWpoint}
\mathcal{C}^{\bm{W}}=\sum^{2}_{n=0}\mathcal{C}_{n}^{\bm{W}} \approx 0.
\end{split}
\end{equation}
Concerning the $\bm{\Gamma}$ point, in our numerical calculations we obtain that the local Chern number around $\bm{\Gamma}$ is $\mathcal{C}_{0,1}(\bm{\Gamma})\approx\pm 1$ and comes from the numerical result of three points glued together around $\bm{\Gamma}$-point each contributing by a factor of $1/3$. Note that, in the limit $\delta\rightarrow 0$, all the curvature in $\bm{\Gamma}$ is concentrated in a singular point of $\pm2\pi$ flux, and if we integrate around, we obtain a local Chern number of $\pm1$. The sign changes between the central band and the first remote band, while in the case of the second remote band $\mathcal{C}_{2}(\bm{\Gamma})\approx0$. We can summarize these numerical results around the $\bm{\Gamma}$-point as, 
\begin{equation}
\begin{split}\label{eq:berryGammapoint}
&\mathcal{C}_{0}(\bm{\Gamma})\approx 3(1/3)=1\\
&\mathcal{C}_{1}(\bm{\Gamma})\approx-3(1/3)=-1\\
&\mathcal{C}_{2}(\bm{\Gamma})\approx0.\\
\end{split}
\end{equation}
Therefore, there is also a sum rule between the three bands equivalent to Eq. (\ref{eq:berryWpoint}),
\begin{equation}
\begin{split}\label{eq:berryGammapointA}
\mathcal{C}^{\bm{\Gamma}}=\sum^{2}_{n=0}\mathcal{C}_{n}(\bm{\Gamma}) \approx 0
\end{split}
\end{equation}
where $n$ runs over the band number. Here $\mathcal{C}_{0}(\bm{\Gamma})$, $\mathcal{C}_{1}(\bm{\Gamma})$, and $\mathcal{C}_{2}(\bm{\Gamma})$ are the Berry fluxes in the central band, first remote band and second remote band, respectively. 
This balance between curvatures comes from an interchange between the upper middle band and the two first remote bands as shown in Fig.~\ref{fig:alpha1} and  Fig.~\ref{fig:chern_vs_gap}. Such conservation rules at hybridization points are reminiscent of the Chern meeting formula analytically deduced for the Quantum Hall effect when two topologically distinct phases merge~\cite{Naumismap_2016}. Such meeting phases produce Van Hove singularities that preserve the topological information of their parenting phases~\cite{Naumismap_2016,Naumis_QCandQHM_2019}. 

\begin{figure}[tb]
\includegraphics[scale=0.55]{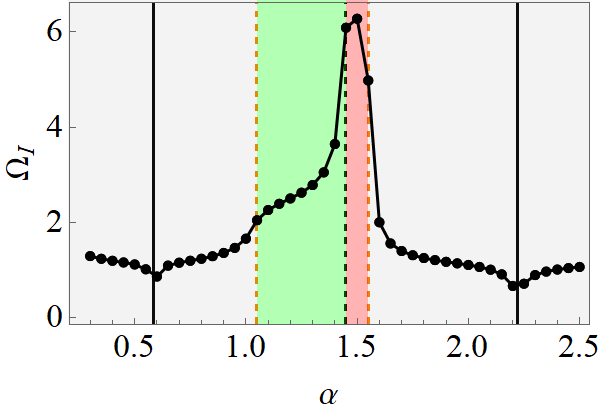}
\caption{The Marzari-Vanderbilt cummulant defined as $\Omega_{I}=\frac{1}{N_c}\sum_n\sum_{\bm{k}}Tr(g^n(\bm{k}))$ as a function of $\alpha$. We show the results for the first hybridization region. The quantum metric reaches its maximum value when the central band touches higher bands. More importantly, the gap closing at $\bm{M}$ (red region) leads to a higher increase in the quantum metric than the gap closing at $\bm{\Gamma}$ (green region). Magic angles are indicated here with black solid vertical lines. $\Omega_{I}$ is minimal at magic angles. For these numerical results, we fixed $n=2$.}
\label{fig:cumulantt}
\end{figure}

Another important aspect of topology is the quantum metric distribution~\cite{kukka2022}. It allows further characterization of the bands and is important in correlated effects such as superconductivity~\cite{Long2024Evolution}. In the case of flat bands, this relationship between correlations and quantum metric (also called Fubini-Study metric) is even more important; see for example~\cite{2023Jin-Xin,2021ROSSI,Manato2021}.

A useful related mathematical tool to quantify the degree of the wave function inhomogeneity is the Marzari-Vanderbilt cumulant, defined as \cite{Vanderbilt1997}, 

\begin{equation}
\begin{split}\label{eq:OMEGA_I}
\Omega_I&=\frac{A_{0}}{(2\pi)^2}\sum_n\int_{BZ}Tr(g^n(\bm{k}))d\bm{k}\\
&=\frac{1}{N_{c}}\sum_n\sum_{\bm{k}}Tr(g^n(\bm{k})), 
\end{split}
\end{equation}
where 
\begin{equation}
\begin{split}\label{eq:trace}
Tr (g^{n}(\bm{k}))= g_{xx}^n(\bm{k})+g_{yy}^n(\bm{k})
\end{split}
\end{equation}
is the trace of the quantum metric tensor, which is detailed in the appendix \ref{Sec: Marzari-Vanderbilt}. $N_c$ is the number of $\bm{k}$ points in the numerical grid, BZ denotes the Brillouin zone, and $A_0=8\pi^2 N/3\sqrt{3}$ is the area of the moiré unit cell with $N$ the real-space unit cells in the system. 

The Marzari-Vanderbilt (MV) parameter measures the similarity of the states in one or in composite bands. In addition, as seen in the appendix \ref{Sec: Marzari-Vanderbilt}, the overlap of states $\langle\Psi_{n}(\bm{k})|\Psi_{m}(\bm{k}+\bm{q})\rangle$ weights the coupling process, and therefore, this overlap measures the similarity of $|\Psi_{m}(\bm{k}+\bm{q})\rangle$ and $|\Psi_{n}(\bm{k})\rangle$. Here $\bm{q}$ is a small change in momentum $\bm{k}$, i.e., $\bm{q}=\Delta \bm{k}\rightarrow0$. 

Panels in Fig.~\ref{fig:berry}c) and Fig.~\ref{fig:berry}d) show the density distribution of $\text{Tr}(g^{n}(\bm{k}))$. Small values indicate that the wavefunctions around a given $\bm{k}$ point are similar, while large values reflect significant variations in their charge distribution. Notably, despite these being dispersive bands, there are extended regions where $\text{Tr}(g^{n}(\bm{k}))$ is minimal (shown in blue). The most delocalized wavefunctions appear near the transition points, where $\text{Tr}(g^{n}(\bm{k}))$ reaches its maximum. Unlike flat bands, dispersive bands contribute to superfluid stiffness through both conventional and geometric terms. In contrast, for flat bands, the conventional stiffness vanishes~\cite{kukka2022}.

An interesting effect is the evolution of the $\Omega_I$ parameter with the twist angle. Figure~\ref{fig:cumulantt} shows how $\Omega_I$ varies as a function of the $\alpha$ parameter. Notably, near the magic angles (indicated by black vertical lines), $\Omega_I$ reaches minimum values. This behavior arises because the wavefunctions near the magic angles are very similar to one another. As previously discussed, except at the $\Gamma$ point—where the charge distribution forms a ring around the AA centers—the charge at other high-symmetry points is localized at the AA centers (see also Fig.~\ref{fig:alpha1})\cite{Guinea2018Electrostatic,Rademaker2019Charge}. This strong localization is responsible for the two minima in $\Omega_I$ observed at the magic angles.

On the other hand, during the transition from phase $P_1$ to $P_2$ (associated with the $\Gamma$ point), represented by the gray-to-green region in Fig.~\ref{fig:cumulantt}, $\Omega_I$ shows a slight increase, which then rises sharply and reaches a maximum at the transition from phase $P_2$ to $P_4$ (associated with the $M$ point). After this transition, the curve decreases and approaches a second minimum at the second magic angle. Notably, the peak observed during the transition at the $M$ point is consistent with the delocalization process revealed by the inverse participation ratio as a function of the twist angle reported in Ref.~\cite{Naumis2022}.

\section{Non-abelian magnetic energy}\label{Sec:Non-Abelian}
 
 As shown in the appendix \ref{sec:squared}, the squared Hamiltonian can be written as, 
\begin{equation}
\begin{split}
H^2 &=(-\laplacian+\bm{A}^{2})\tau_0+i\alpha^{2}[A_{x}, A_{y}]\tau_z-2i\alpha\hat{\bm{A}}\cdot\gradient \\
&+\alpha(\partial_{x}\hat{A}_y-\partial_{y}\hat{A}_x)
    \label{eq:H22}
\end{split}
\end{equation}
where $\hat{\tau}_{j}$ (with $j=1,2,3$) is the set of Pauli matrices in the pseudo-spin layer degree, and the identity $2 \times 2$ matrix $\hat{\tau}_{0}$. $A_x$ and $A_y$ are defined in the appendix \ref{sec:squared}. Its SU(2) matrix versions $\hat{A}_x$ and $\hat{A}_y$ are, 

\begin{equation}
\begin{split}
\hat{\bm{A}}_{x}=\bm{A}_{1,x}\tau_1+\bm{A}_{2,x}\tau_2  \\
\hat{\bm{A}}_{y}=\bm{A}_{1,y}\tau_1+\bm{A}_{2,y}\tau_2
    \label{eq:components_A}
\end{split}
\end{equation}
Explicitly, the components of $\hat{\bm{A}}$ are,
\begin{equation}
\begin{split}
\bm{A}_{1,\nu}&=\sum_{\mu}\cos{(\bm{q}_{\mu}\cdot\bm{r})}\bm{q}_{\mu}^{\perp,\nu},\\
\bm{A}_{2,\nu}&=\sum_{\mu}\sin{(\bm{q}_{\mu}\cdot\bm{r})}\bm{q}_{\mu}^{\perp,\nu},\\.
  \end{split} 
  \end{equation}
where $\nu=x,y$ is the index for the spatial coordinates. Note that $\hat{\bm{A}}$ is non-Abelian as follows from the fact that $[\hat{\bm{A}}_{\mu}(\bm{r}),\hat{\bm{A}}_{\nu}(\bm{r}^{\prime})]\neq 0$, i.e.,  it does not commute with itself at different locations.  Written in such a way, we can identify the Zeeman coupling energy,
\begin{equation}
\begin{split}
\hat{F}_{xy} &= \partial_{x}\hat{A}_y-\partial_{y}\hat{A}_x +i\alpha[\hat{A}_x,\hat{A}_y]\\
&= -\hat{\bm{B}}\cdot\hat{\bm{\tau}}+i\alpha[\hat{A}_x,\hat{A}_y]
\label{eq:Zeeman_coupling}
\end{split}
\end{equation}
where upper hats represent matrices. For convenience, we re-scale the spatial coordinates as $\bm{r} \rightarrow \bm{r}/\alpha$ from where $\gradient \rightarrow \alpha\gradient$, resulting in a re-scaled position Hamiltonian $H_s=H/\alpha$, 

\begin{equation}
\begin{split}
H_s^2&=(-\laplacian+\bm{A}^{2}(\bm{r}/\alpha))\tau_0+i[A_{x}(\bm{r}/\alpha), A_{y}(\bm{r}/\alpha)]\tau_z \\
& -2i\hat{\bm{A}}(\bm{r}/\alpha)\cdot\gradient-\frac{1}{\alpha}\hat{\bm{B}}(\bm{r}/\alpha)\cdot\hat{\bm{\tau}}.
    \label{eq:H2_scaled}
\end{split}
\end{equation}

\begin{figure*}[tb]
\centering
{
\includegraphics[scale=0.35]{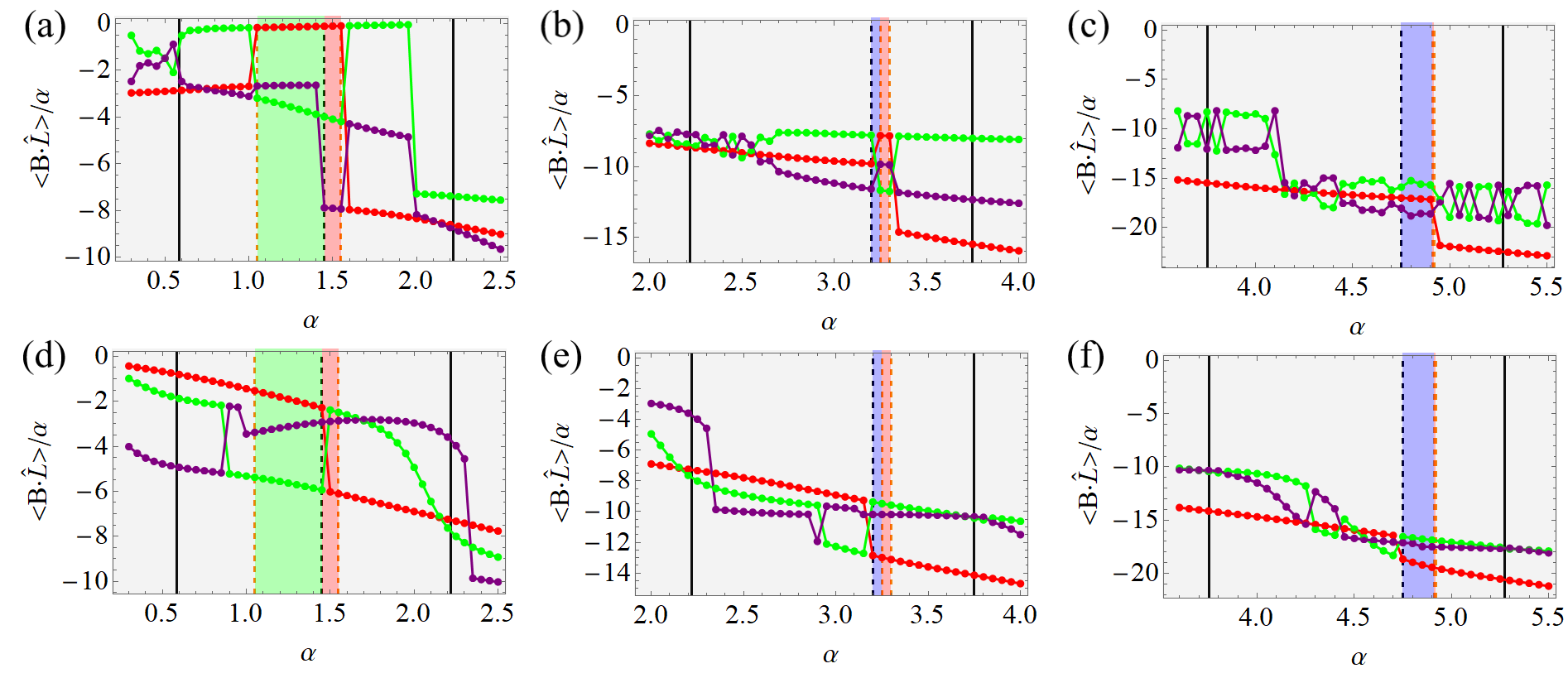}
}
\caption{Scaled magnetic energies $\langle\bm{B}\cdot\hat{\bm{L}}\rangle/\alpha$ as a function of $\alpha$. Magnetic energies at, (a)-(c) the $\bm{\Gamma}$ point, and (d)-(f) the $\bm{M}$ point. The central band (red), first remote band (green), and second remote band (purple) are indicated. There is a clear interplay between bands in the hybridization regions. Magic angles (Black vertical lines) and topological phase transitions (black and orange dashed lines) are indicated.}
\label{fig:energiesBL} 
\end{figure*}

The strong confinement of electrons allows us to suppose an almost uniform magnetic field. Therefore, we can write $\bm{A}\cdot\hat{\bm{p}}\approx -\bm{B}\cdot\hat{\bm{L}}$ where $\hat{\bm{L}}$ is the total angular moment \cite{NN2024}.  Under such simplification, the re-scaled Hamiltonian is, 

\begin{equation}
\begin{split}
\hat{H_{s}}^2&=(-\laplacian+\bm{A}^{2}(\bm{r}/\alpha))\tau_0+i[A_{x}(\bm{r}/\alpha), A_{y}(\bm{r}/\alpha)]\tau_z\\
&-\hat{\bm{B}}(\bm{r}/\alpha)\cdot(2\hat{\bm{L}}+\frac{\hat{\bm{\tau}}}{\alpha})
\label{eq:H2_Magnetic_off_diagonal2}
\end{split}
\end{equation}

The previous expression allows to identify a Zeeman contribution $E_{Zeeman}=\langle \hat{\bm{B}}\cdot\hat{\bm{\tau}} \rangle$ where $\hat{\bm{\tau}}=(\tau_x,\tau_y)$ are the Pauli matrices in the layer degree of freedom.
Note that only the last term depends on $\alpha$ and taking the asymptotic limit $\alpha\rightarrow\infty$ we have that the Zeeman energy $-\frac{1}{\alpha}\hat{\bm{B}}(\bm{r}/\alpha)\cdot\hat{\bm{\tau}}\rightarrow 0$. Therefore, it is expected to be similarly small for higher magic angles, and neglecting it should not significantly impact the results. 

We can also define a magnetic energy $E_{Magnetic}=-2\langle\hat{\bm{B}}\cdot\hat{\bm{L}}\rangle$ which is one of the off-diagonal terms in the square Hamiltonian. In the asymptotic limit it is easy to see that $E_{Magnetic}>>E_{Zeeman}$.  Hence, the Hamiltonian in this limit can be simplified into,
\begin{equation}
H_s^2=(\bm{p}+\hat{\bm{A}}(\bm{r}/\alpha))^{2}+i[A_{x}(\bm{r}/\alpha), A_{y}(\bm{r}/\alpha)]\tau_z
    \label{eq:H2_scaled_limit}
\end{equation}
where $\hat{H}^2=(H/\alpha)^2$ and $\bm{p}=-i\gradient$ is the canonical momentum operator. Accordingly, $\hat{H}^2$ it's expected to have a non-Abelian QHE.

Here, we are more interested in the first hybridization region. Therefore, understanding how the magnetic energy is intertwined between the lowest energy bands is important. For that reason, in Fig.~\ref{fig:energiesBL}, the expected valie fo the orbital magnetic energy per layer $\langle \Psi_{\bm{k}}(\bm{r})|\hat{\bm{B}}\cdot\bm{\hat{L}} | \Psi_{\bm{k}}(\bm{r}) \rangle/\alpha$ is plotted as a function of $\alpha$ for the first three hybridization regions. The orbital magnetic energy has an interesting behavior. For instance, in Fig.~\ref{fig:energiesBL}(a)-(c), the magnetic energy at the $\bm{\Gamma}$ point shows clear evidence that the touching of bands generates an orbital energy interchange between bands. Importantly, the first and second remote bands fluctuate intrinsically between their magnetic energies from the hybridization region. On the other hand,  in Fig.~\ref{fig:energiesBL}(d)-(f) corresponding to the $\bm{M}$ point, there are jumps in the orbital magnetic energy at the topological phase transition where the central band touches the first remote band. In $\bm{M}$, the jumps are not exactly at the hybridization points because the Zeeman energy is finite and is creating this shifting difference \cite{NN2024}. However, at $\bm{\Gamma}$ the effect of the Zeeman term is minimal.\\ 

\section{Conclusion}\label{secConclusion}

New topological phase transitions and Chern insulating phases with Chern number $C = \pm 2$ in the central bands of twisted bilayer graphene (TBG) have been identified for twist angles between the first and second magic angles. These phases arise from the hybridization between the central flat band and the first and second remote bands. Whenever these remote bands touch the central one, a topological transition occurs. These transitions are governed by band inversions at high-symmetry points in reciprocal space, specifically at $\bm{\Gamma}$ and $\bm{M}$. The sequence of topological phases that appears between the first and second magic angles is different from the one observed at higher twist angles. This difference is associated with a transition from a non-Abelian to an Abelian effective field description~\cite{Naumis2023r}.

To further analyze the topology, we considered the quantum geometry through the quantum metric. A clear relationship was found between the band inversions and the redistribution of charge at $\bm{\Gamma}$ and $\bm{M}$. Both the Berry curvature and the quantum metric exhibit sharp features localized around these points, highlighting the system’s sensitivity to hybridization between the central and remote bands.

Interestingly, we also observed a reorganization of the charge distribution at the $\bm{K}$ and $\bm{K}'$ points, even though these Dirac points are not directly involved in the gap closings. This suggests the presence of nonlocal effects, where a band inversion at $\bm{\Gamma}$ or $\bm{M}$ can influence the electronic states throughout the mini Brillouin zone. In particular, the charge near the $\bm{K}$ points evolves from being concentrated at the AA centers to forming ring-like puddles around them, indicating a redistribution driven by the global modification of the moiré potential.

The Marzari–Vanderbilt cumulant reaches its maximum in the regions where the central band touches higher-energy bands, indicating enhanced wavefunction delocalization \cite{Mizoguchi_2019}. In particular, the transition from topological phase $P_2$ to phase $P_3$, which occurs around $\alpha \approx 1.45$ to $\alpha \approx 1.55$, produces a pronounced peak in $\Omega_I$. We encourage experimental exploration of this regime, for example, by applying an interlayer potential of approximately $V_{\delta} \approx 10$ meV.

Furthermore, we observed a reorganization of the orbital magnetic moments involving the central, first, and second bands during hybridization. This redistribution occurs at the $\bm{\Gamma}$ and $\bm{M}$ points and reflects changes in the orbital character of the bands across the transitions, revealing a rich topological structure.

Although the chiral model is an approximation, it becomes more accurate at small twist angles due to stronger lattice relaxation effects \cite{NN2024}. As a result, the new topological phases identified in the first hybridization region may be experimentally accessible. The emergence of $C = \pm 2$ Chern numbers in TBG, which had not been reported previously, is a significant finding. Such high Chern numbers, when combined with electronic correlations, could give rise to strongly correlated phases such as fractional Chern insulators \cite{Jarrillo2021_Fractional, Ledwidth2023, 2022Ledwith_Vortex, Ledwidth2020}. Therefore, investigating this topological region is essential for the realization of non-Abelian phases of matter~\cite{2025Ledwith}, with potential applications in quantum computing and quantum information technologies~\cite{2024NL,NL2025}.

\section*{Acknowledgements}
We thank Zhen Zhan, Alejandro Jimeno-Pozo and Danna Liu for discussions.\ This work was supported by  CONAHCyT project 1564464 and UNAM DGAPA project IN101924.\ Leonardo Navarro is supported by a CONAHCyT PhD scholarship No.\ 1564464.\ P.A.P and F.G acknowledges support from NOVMOMAT, project PID2022-142162NB-I00 funded by MICIU/AEI/10.13039/501100011033 and by FEDER, UE as well as financial support through the (MAD2D-CM)-MRR MATERIALES AVANZADOS-IMDEA-NC.\ F.G acknowledges the support from the  Department of Education of the Basque Government through the project \texttt{PIBA\textbackslash 2023\textbackslash 1\textbackslash 0007(STRAINER)}.

 \section{Appendix}\label{secAppendix}

\subsection{Squared Hamiltonian}\label{sec:squared}

In our previous works \cite{Naumis2021, Naumis2022, Naumis2023,NN2024,Navarro-Labastida_2025}, we demonstrated that squaring the Hamiltonian $\mathcal{H}$ allows us to simplify it into a $2 \times 2$ matrix that we call the squared Hamiltonian  $H^{2}$,
\begin{equation}\label{eq:H2}
\begin{split}
&H^{2}=\\
&\begin{pmatrix} -\nabla^{2}+\alpha^{2}(\bm{A}^{2}+\Delta)&  \alpha (-2i\bm{A}\cdot\nabla +\bm{B}) \\
 \alpha (-2i\bm{A}^{*}\cdot\nabla +\bm{B}^{*}) & -\nabla^{2}+\alpha^{2}(\bm{A}^{2}-\Delta) 
  \end{pmatrix} 
  \end{split} 
  \end{equation}
where we defined, 
\begin{equation}
 \begin{split}
\bm{A}&=\sum_{\nu=1}^{3}e^{-i\bm{q}_{\nu}\cdot\bm{r}}\bm{q}_{\nu}^{\perp}.\\
  \end{split} 
\end{equation}
$\bm{A}$ is a pseudo-magnetic vector potential with $C_3$ symmetry and $\bm{A}^{2}=|\bm{A}|^{2}$.  The squared norm of the coupling potential is an effective intralayer confinement potential,
\begin{equation}
\begin{split}
|U(\pm\bm{r})|^{2} &= \bm{A}^{2}\mp \Delta
\end{split}
\end{equation}
where the confinement potential $|U(\pm\bm{r})|^{2}$ is separated into its purely symmetric $\bm{A}^2(\bm{r})$ and anti-symmetric $\Delta(\bm{r})$ parts defined as, 

\begin{equation}
\begin{split}
\bm{A}^2(\bm{r})= 3-\sum_{\nu=1}^{3}\cos{(\bm{b}_{\nu}\cdot\bm{r})}\\
\Delta(\bm{r})=\sqrt{3}\sum_{\nu}(-1)^{\nu}\sin{(\bm{b}_{\nu}\cdot\bm{r})}
\label{eq:potentials_s_anti}
\end{split}
\end{equation}
where,
\begin{equation}
    \Delta(\bm{r})=i[A_x,A_y].
\end{equation}
$A_x$ and $A_y$ are the non-Abelian components of the $SU(2)$ pseudo-magnetic vector potential (See Appendix A). It is important to remark that the pseudo-magnetic vector potential satisfies the relation $\gradient\cdot\bm{A}=0$, and thus is a Coulomb gauge invariant field and $\gradient\times\bm{A}=\bm{B}$ (layer 1) and $\gradient\times\bm{A}^{*}=\bm{B}^{*}$  (layer 2).
The magnetic field is given by,
\begin{equation}
    \bm{B}=-i \sum_{\nu=1}^{3}e^{-i\bm{q}_{\nu}\cdot\bm{r}}\bm{e}_z
\end{equation}
where we have used the identity $\bm{e}_z=\bm{q}_{\nu}\times\bm{q}^{\perp}_{\nu}$ and $\bm{e}_z$ is a unit vector in the direction perpendicular to the graphene's plane. The squaring of the chiral TBG model is akin to a supersymmetric transformation \cite{DikiMatsumoto2023, Hatsugai2020, TomonariMizoguchi2022, Yoshida2021,2023Mizoguchi}.\\

\subsection{Quantum Metric: Marzari-Vanderbilt} \label{Sec: Marzari-Vanderbilt}

For clarity, in this appendix we drop any band indexes attached to the wavefunctions, later on they will be restored. Based on the infinitesimal Bures distance \cite{2024Abdiel},
\begin{equation}
\begin{split}\label{eq:bures}
D^{2}_{12}=1-|\langle\partial_{\mu}\Psi(\bm{k})|\Psi(\bm{k}+d\bm{k})\rangle|^{2}
\end{split}
\end{equation}
and using the Taylor expansion in the infinitesimal wavefunction,

\begin{equation}
\begin{split}\label{eq:taylor}
|\Psi(\bm{k}+d\bm{k})\rangle &\approx |\Psi(\bm{k}))\rangle+ \sum_{\alpha}|\partial_{\alpha}\Psi(\bm{k})\rangle d\bm{k}_{\alpha}\\
&+ \frac{1}{2}\sum_{\alpha,\beta}|\partial_{\beta}\partial_{\alpha}\Psi(\bm{k})\rangle d\bm{k}_{\alpha}d\bm{k}_{\beta}
\end{split}
\end{equation}
introducing Eq. (\ref{eq:taylor}) into Eq. (\ref{eq:bures}) and keeping terms up to the second order, it follows that,

\begin{equation}
\begin{split}\label{eq:Bure_metric}
D^{2}_{\bm{k},\bm{k}+d\bm{k}}=\sum_{\alpha,\beta}g_{\alpha,\beta}(\bm{k})d\bm{k}_{\alpha}d\bm{k}_{\beta}
\end{split}
\end{equation}
where we have the quantum metric tensor, also called the Fubini-Study metric \cite{2024Abdiel}, 
\begin{equation}
\begin{split}\label{eq:fubini-study}
g_{\alpha,\beta}(\bm{k})&=Re(\langle\partial_{\alpha}\Psi(\bm{k})|\partial_{\beta}\Psi(\bm{k})\rangle\\
&-\langle\partial_{\alpha}\Psi(\bm{k})|\Psi(\bm{k})\rangle\langle\Psi(\bm{k})|\partial_{\beta}\Psi(\bm{k})\rangle)
\end{split}
\end{equation}
By using the projector
\begin{equation}
\begin{split}\label{eq:Qmu}
Q_{\bm{k}}=1-|\Psi(\bm{k})\rangle\langle\Psi(\bm{k})|
\end{split}
\end{equation}

the quantum geometric tensor is re-expressed as, 
\begin{equation}
\begin{split}\label{eq:gmunu}
\bm{g}_{\alpha,\beta}(\bm{k})=Re(\langle\partial_{\alpha}\Psi(\bm{k})|Q_{\bm{k}}|\partial_{\beta}\Psi(\bm{k})\rangle)
\end{split}
\end{equation}
where $\partial_{\alpha}=\partial/\partial_{\bm{k}_{\alpha}}$.

In calculating the matrix tensor, we take the difference between one wavefunction and its neighbors, through the projection operator $Q_{\bm{k}}=1-|\Psi(\bm{k})\rangle\langle\Psi(\bm{k})|$, this difference is projected into the orthogonal subspace of $|\Psi(\bm{k})\rangle$. Then the sum of the trace of the metric tensor measures the intrinsic homogeneity of the underlying Hilbert space and the trace of the metric tensor at each value of $\bm{k}$ measures the degree of mismatch between the neighboring Bloch subspaces $\bm{k}$ and $\bm{k} + \bm{q}$, for $\bm{q}$ a small change in momentum $\bm{k}$, i.e., $\bm{q}=\Delta \bm{k}\rightarrow0$.


\end{document}